\documentclass[journal=nalefd,manuscript=letter]{achemso}

\usepackage[T1]{fontenc} 
\usepackage[version=4]{mhchem}

\usepackage{siunitx}
\usepackage{physics}
\author{Dasika Shishir}
\affiliation{Department of Electrical Engineering, Indian Institute of Technology Bombay, Powai, Mumbai-400076, India.}
\email{dshishir@iitb.ac.in}

\author{Madhur Parashar}
\affiliation{School of Medical Science and Technology, Indian Institute of Technology Kharagpur, Kharagpur, West Bengal, 721302, India}
\email{madhurparashar@iitkgp.ac.in}

\author{Kasturi Saha}
\affiliation{Department of Electrical Engineering, Indian Institute of Technology Bombay, Powai, Mumbai-400076, India.}
\email{kasturis@ee.iitb.ac.in}
\title{Mapping AC Susceptibility with Quantum Diamond Microscope}

\abbreviations{NMR,NV}

\begin{document}
\begin{abstract}
We present a novel technique for determining the microscale AC susceptibility of magnetic materials. We use
magnetic field sensing properties of nitrogen-vacancy (\ce{NV-}) centers in diamond to gather quantitative 
data about the magnetic state of the magnetic material under investigation. In order to achieve the
requisite speed in imaging, a lock-in camera is used to perform pixel-by-pixel lock-in detection of \ce{NV-} 
photo-luminescence. In addition, a
secondary sensor is employed to isolate the effect of the excitation field from fields arising from magnetic 
structures 
on \ce{NV-} centers. We demonstrate our experimental technique by measuring the AC susceptibility of soft 
permalloy micro-magnets at excitation frequencies of up to \SI{20}{\hertz} with a spatial resolution of \SI{1.2}{\micro \meter} and a field of view of \SI{100}{\um}. Our work paves the way for microscopic measurement of AC susceptibilities of magnetic materials relevant to physical, biological, and material sciences. 
\end{abstract}
    
\section{Introduction}

Measurement of AC susceptibility is critical and indispensable in investigating magnetic relaxations and 
phase transitions in various magnetic materials\,\cite{Topping_2018,balanda2013ac}. In AC susceptibility 
measurements, a small external 
perturbation of frequency ranging from a few hertz to several tens
of kilohertz is applied to observe the dynamic change in the magnetic moment. The magnetic system's relaxation 
time under investigation, $\tau$, dictates the system's frequency response. 
Hence, AC susceptibility measurements lead to the characterization of relaxation times. AC 
susceptibility measurements have been performed in the context of high critical temperature 
superconductors\,\cite{PhysRevB.41.8937,OCKO2004231}, spin-glasses,  paramagnetic 
salts\,\cite{PhysRevLett.120.087201,PhysRevB.98.144436}, magnetic 
nano-particles\,\cite{doi:10.1063/1.5120609,Zhong_2017,Wu2019},and 
conventional and low-dimensional ferromagnets\,\cite{PhysRevB.82.064404,PRXQuantum.2.030352}.
In conventional AC susceptometry, the sample under study is excited using a primary coil. The in-phase and 
quadrature-phase change in magnetization is detected through a secondary coil 
using a lock-in amplifier\,\cite{Laurent_2008}. The quadrature-phase signal obtained is proportional to 
the real part of the susceptibility $\chi'$, and the in-phase signal is proportional
to the imaginary part of the susceptibility $\chi''$. In general, any magnetometer can be turned 
into a susceptometer if the frequency response of the magnetometer is adequate to measure the change in 
magnetization at the desired excitation frequencies. As such, SQUID 
magnetometers\,\cite{doi:10.1063/1.1144664,doi:10.1063/1.5045299}, MOKE 
magnetometers\,\cite{PhysRevB.101.134404,doi:10.1063/1.1148368}, and electron, neutron, and X-Ray 
based methods have\,\cite{doi:10.1063/1.3360205,ASPELMEIER1995256} been utilized to measure AC susceptibility. However, some of these techniques  do not give 
information on 
the local microscale variations in susceptibility, which might be of great value in many cases, say near the 
vortices in a high-Tc superconductor. Some of the methods described above, though they have good spatial 
resolution down to \SI{10}{\nm}, fail to provide quantitative information on magnetic properties like
dipole moment and susceptibility. 

In this letter, we present a spatially resolved dynamic AC susceptibility experimental approach and illustrate it by measuring the AC susceptibility of a collection of microfabricated permalloy 
(\ce{Ni_{0.8}Fe_{0.2}}) micromagnets. This work builds upon the recent
demonstrations of high-speed wide-field magnetic imaging using a per-pixel lock-in amplifier 
camera\,\cite{Parashar2022,PhysRevApplied.17.064051} in
a quantum diamond microscope\,\cite{doi:10.1063/5.0066733}. A lock-in camera significantly enhances the
per-pixel sensitivity due to the rejection of noise outside the modulation frequency bandwidth. 
Additionally, a lock-in camera utilizes the full-well capacity of a CMOS sensor more efficiently to enhance the 
dynamic range of the 
collected fluorescence\,\cite{doi:10.1063/1.5010282,Parashar2022}. To perform dynamic magnetic imaging, 
a single microwave frequency is chosen in the spectrum of negatively 
charged nitrogen-vacancy (\ce{NV-}) centers, where the photoluminescence (PL) varies linearly with
the change in external magnetic field\,\cite{doi:10.1073/pnas.1601513113,Clevenson2015}. The linear 
range is limited to the linewidth of the spectrum, which is typically \SIrange{1}{2}{\mega \hertz}. This
frequency shift corresponds to a change in the magnetic field of \SIrange{10}{20}{\micro \tesla}. However, the AC excitation field is varied by a few 
milli-Teslas\,\cite{balanda2013ac} to perform AC susceptibility measurements. This variation falls outside the dynamic range of lockin-camera-based
quantum diamond microscope, limiting its use in dynamic imaging of magnetic materials. To mitigate the
problem mentioned above, we use a secondary sensor to dynamically set the microwave frequency such that  
\ce{NV-} PL response falls in a regime that is always linear to the changes in small magnetic fields 
arising from the magnetic material under study. Using a second sensor effectively decouples the 
excitation field from the fields arising from the magnetic materials as a response to the excitation field.
With this technique, a variety of excitation fields may be used to investigate different materials with reasonable SNR. For instance, a magnetic material with low susceptibility that needs a strong excitation 
to cause a change in magnetic moment can also be detected. Using
our technique, we detect a change in the magnetic moment of \SI{36}{\femto \joule \per \tesla} by applying an
AC  field of \SI{1}{\milli \tesla} of up-to \SI{20}{\hertz} while maintaining a spatial resolution of \SI{1.2}{\micro \meter}. 
\section{Results and Discussion} 

\begin{figure}[ht]
  \centering 
  \includegraphics[width=\textwidth]{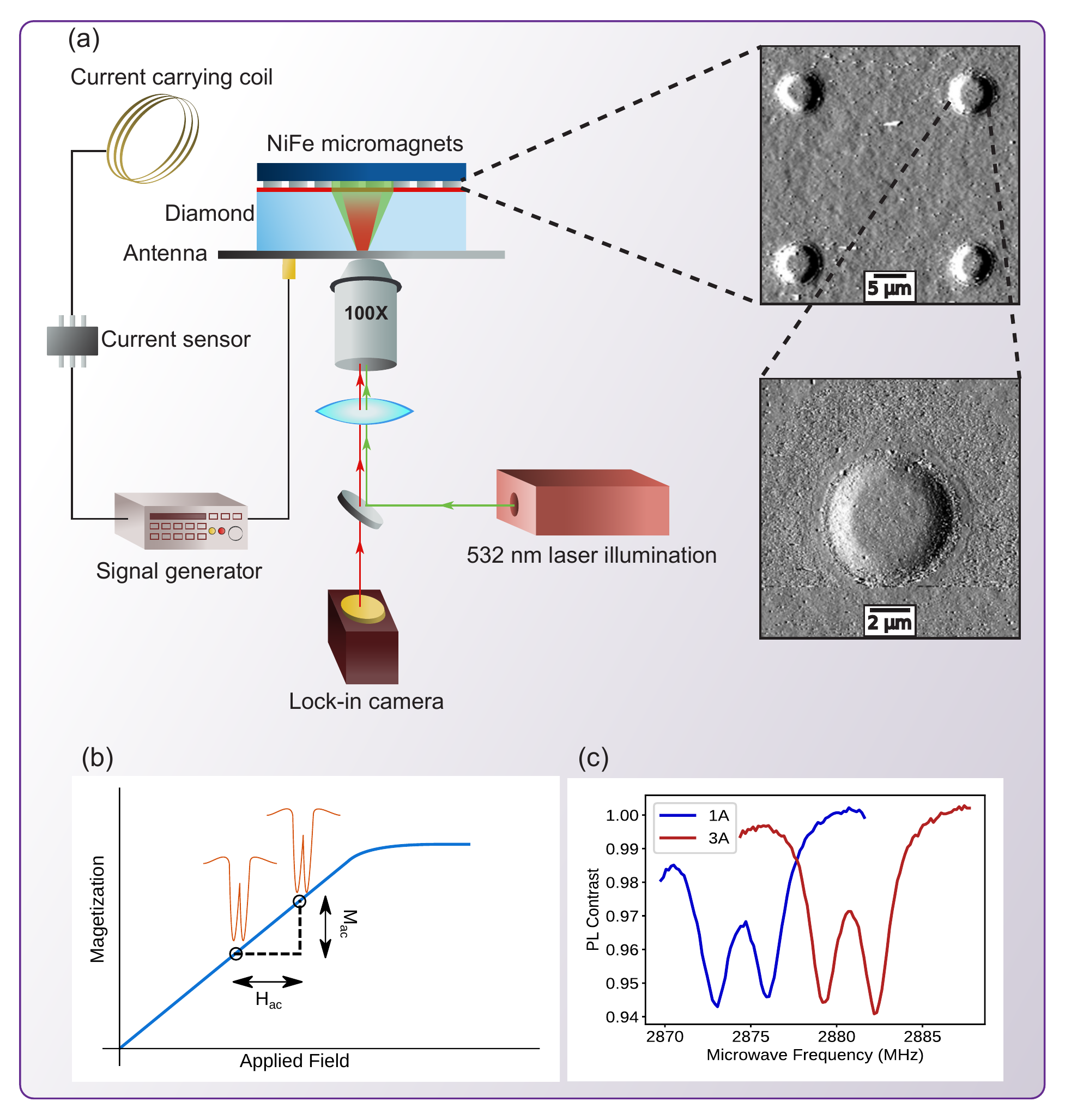}
  \caption{(a). A schematic representation of the fabricated \ce{Py} micro-magnet array placed on top of a
  \SI{1}{\micro \meter} thin \ce{NV-} layer in a diamond for wide-field magnetic imaging. A current sensor 
  is used to dynamically set the microwave frequency based on the current passing through the current carrying coil. The inset shows the AFM image of the fabricated \ce{Py} micro-magnets. (b). The stray field from the micro-magnets is measured from ODMR signal at various points
  on the $MH$ curve. The dipole moment can be extracted from the stray field. (c). ODMR signal 
  from a single pixel at two different excitation currents.}
  \label{fig:schem}
\end{figure}

The experimental platform is based on optically detected magnetic resonance (ODMR)\,\cite{Rondin_2014}  signal from \ce{NV-}
centers in diamond is shown in Fig.\,\ref{fig:schem}\,(a).
We place an array of circular Permalloy (\ce{Py}) micromagnets of \SI{5}{\micro \meter} diameter
and \SI{30}{\nano \meter} thickness separated by \SI{25}{\micro\meter} from center to center fabricated on 
a silicon substrate on top of a diamond chip. The diamond chip is $4\times4\,$\SI{}{\milli\meter ^2} 
electronic grade,  $99.99\,$\% \ce{^{12}_{}C} diamond from Element 6, with approximately \SI{1}{\micro 
\meter} thick 
\ce{^{15}_{}N} doping. The concentration of \ce{NV-} in this layer is around 1-2\,ppm. The diamond crystal 
has $\qty{100}$ front facet and ⟨110⟩ edge orientation.
 A \SI{532}{\nm}, \SI{.5}{\watt} laser is incident on the diamond layer
through a $100\,\times$ air objective of numerical aperture \num{0.9}. The overall magnification of 
the system is $34\times$. Each camera pixel is \SI{40}{\um }. Hence, each pixel captures PL from 
\SI{1.17}{\um}$\times$\SI{1.17}{\um} area of the diamond. The \ce{NV-} center emits a red PL when excited 
with  the \SI{532}{\nm} laser.
The PL is collected through the same objective and focused on a wide-field
lock-in camera (Heliotis Helicam C3) after passing through necessary filters to reject the green excitation at \SI{532}{\nm}. 
The \ce{NV-} center is a spin-1 system with a bright $m_S=0$ state and relatively dark $m_S=\pm 1$ states. In the absence of a magnetic field, $m_S=\pm 1$ states are nearly
degenerate, and are separated from $m_S=0$ by a zero-field splitting of
\SI{2.87}{\giga\hertz}. In the presence of a magnetic field, the degeneracy 
between $m_S=\pm 1$ is lifted, given as\,\cite{Rondin_2014}
\begin{equation}
 \nu_\pm = D \pm \gamma_e B_{NV},
 \label{eq:gyro}
\end{equation}
where $\nu_\pm$ are resonance frequencies corresponding to $m_S=0\rightarrow \pm 1$ 
transition, $D$ is the zero-field splitting \SI{2.87}{\giga \hertz}, $\gamma_e$ is the 
electron gyro-magnetic ratio equal to \SI{28}{\mega \hertz \per \milli \tesla}, and 
$B_{NV}$ is the magnetic field along one of the four possible diamond crystallographic
orientations. When the frequency of an externally applied microwave signal is close to one of the eight possible resonant frequencies, there is a drop in PL. 
An external magnetic field is applied using a current carrying coil of 
\num{100} turns and \SI{7}{\cm} diameter. At the same time, a signal generator sweeps the microwave 
signal via a loop antenna on which the diamond is placed. The microwave signal is 
amplitude modulated at \SI{8}{\kilo \hertz} using a microwave switch, and the resultant
modulations in the PL are demodulated at a pixel level by the lock-in-camera. A more
detailed account  of the experimental setup and the operation of lock-in-camera in this context is 
provided in Ref\,\cite{Parashar2022}.  

\begin{figure}[htbp]
  \centering
  \includegraphics[width=0.8\linewidth]{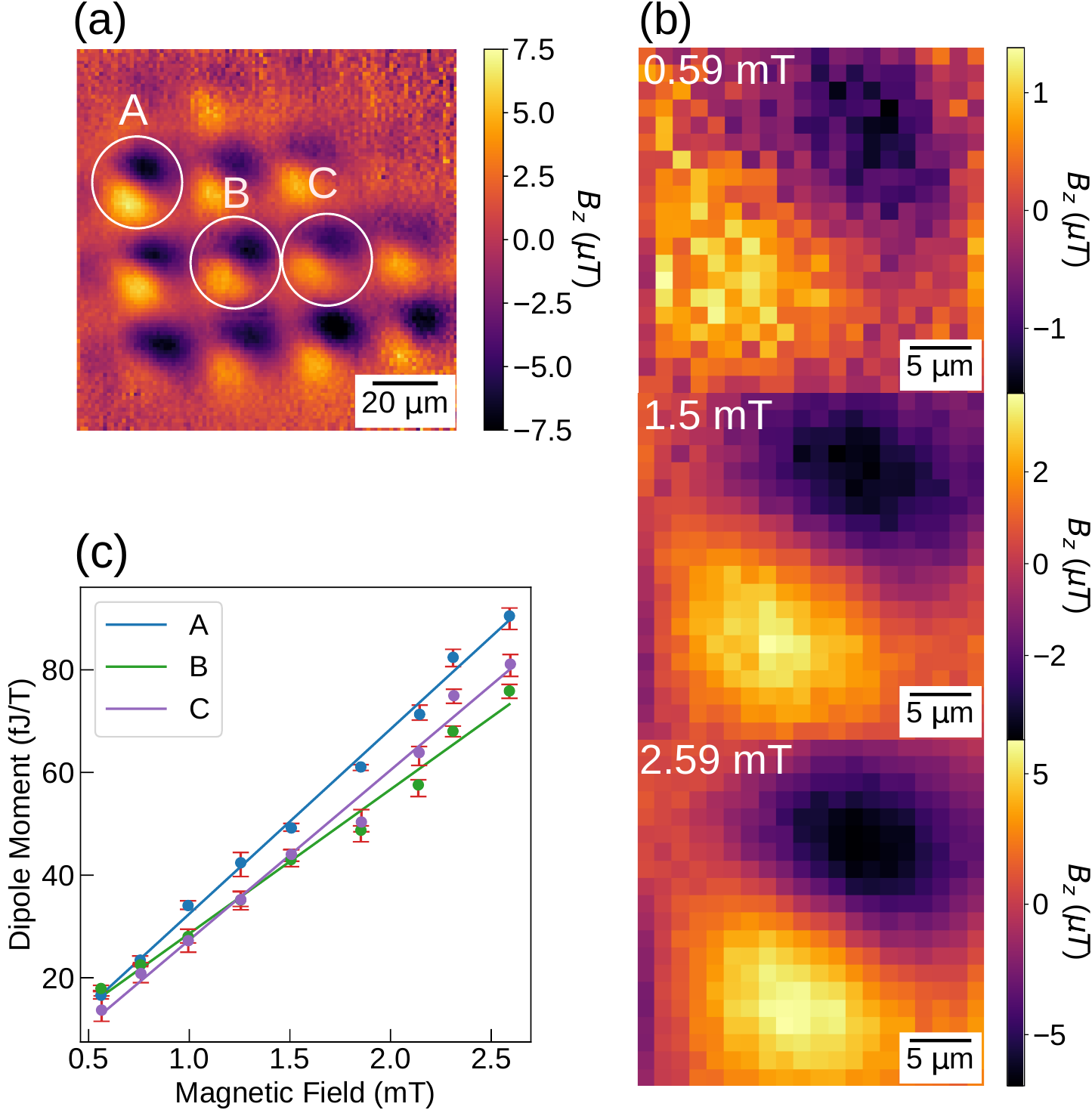}
  \caption{(a). Stray magnetic field along $z$-axis from an array of \ce{NiFe} micro-magnets at an applied field of \SI{2.59}{\milli \tesla}. Three micro-magnets 
  \emph{A}, \emph{B}, and \emph{C} have been shown for reference. (b). Stray magnetic field 
  from micro-magnet \emph{A} at applied fields of \SI{0.56}{\milli \tesla}, \SI{1.5}{\milli \tesla}, and \SI{2.59}{\milli \tesla}. (c). Dipole moment of micro-magnets \emph{A, B, C} plotted against applied magnetic field. Linear functions  $m_A = -3.55 + 36b$, $m_B =  0.5 + 28.1b$, $m_C = -5.76 + 33.11b$ are 
  used to fit the dipole moment of micro-magnets $A, B$, and $C$ versus applied magnetic field. Units
  of $m_A$, $m_B$, $m_C$, and $b$ are same as shown in the plot.}
  \label{fig:dc}
\end{figure}

First, we perform DC characterization of the micro-magnets. To do so, we obtain the full ODMR spectrum
at each value of the applied external field as shown in Fig.\,\ref{fig:schem}(b). The externally applied 
magnetic field not only affects the \ce{NV-} centers in diamond but also affects the 
magnetic state of micro-magnets. In Fig.\ref{fig:dc}, we show the effect of an external
magnetic field on the state of micro-magnets. In Fig.\,\ref{fig:dc}(a), the stray magnetic 
field arising from the the micro-magnetic array is shown. The total applied magnetic field  is
\SI{3.34}{\milli\tesla}, with an in-plane component of \SI{2.59}{\milli \tesla}.  The applied vector 
magnetic fields are obtained by measuring the 
resonance frequencies of all the four possible \ce{NV-} axis and fitting with \ce{NV-} spin 
Hamiltonian\,\cite{PhysRevApplied.10.034044}. Because the shape
anisotropy restricts the dipole moment of the micro-magnet to be in-plane\,\cite{PhysRevB.83.224412}, we 
only consider the applied in-plane magnetic field in susceptibility calculations. 
The stray magnetic field is obtained by extracting
the resonant frequency from each pixel by fitting the ODMR response to 
a function that is a sum of two Lorentzian functions. After the resonant frequency at each pixel is 
obtained, the median resonance frequency is subtracted from all the pixels. The resultant shifts in 
resonant frequencies are converted
to a magnetic field map by dividing the resonant shifts with electron gyro-magnetic ratio as in 
Eq.\,\eqref{eq:gyro}.  
Finally, using the procedure outlined in 
Ref.\cite{doi:10.1063/5.0005335,lima2009obtaining}, and Supplemental Information (Section 3),  the obtained field, which is the field along 
a \ce{NV-} axis is converted into a field along the $z$ axis (pointing out of the diamond 
plane). In Fig.\,\ref{fig:dc}(b), the stray magnetic field from a single micro-magnet at three 
different externally applied magnetic fields - \SI{0.59}{\milli \tesla}, \SI{1.5}{\milli 
\tesla}, and \SI{2.59}{\milli \tesla} are shown. Owing to the soft ferromagnetic nature of
\ce{Py}, we observe that the micro-magnets have very low retentivity. This is 
ascertained by plotting the dipole moment of the micro-magnets as we vary the external magnetic field as 
shown in Fig.\,\ref{fig:dc}(c). The dipole moment is obtained by fitting the magnetic field arising from 
a cylindrical magnet (See Supplemental Information, Section 2). Using the fit, we estimate the stand-off distance between the \ce{NV-} layer and
the micro-magnet layer to be around \SI[parse-numbers = false]{11.51\pm 0.08}{\um}. The dipole moment increases, 
nearly linearly with externally applied field with a slope of \SI{36}{\femto \joule\per \ \tesla \per 
\milli \tesla}, \SI{28.1}{\femto \joule\per \ \tesla \per 
\milli \tesla}, and \SI{33.1}{\femto \joule\per \ \tesla \per 
\milli \tesla} for micromagnets \emph{A}, \emph{B}, \emph{C} as labeled in Fig.\,\ref{fig:dc}(a). Although, we did not perform  measurements on micro-magnets at zero-field, 
by extrapolation, we assume that the dipole moment of
micro-magnets at zero field is very low (\SI{-3.35}{\femto \joule \per \tesla}, \SI{0.5}{\femto \joule \per \tesla}, \SI{-5.76}{\femto \joule \per \tesla} for micromagnets \emph{A}, \emph{B}, \emph{C}) respectively.  With a volume of
\SI{0.59}{\um ^3}, the volume normalized susceptibility of three micro-magnets labeled \emph{A}, \emph{B},
and \emph{C} were found to be \num{69.7}, \num{54.4}, \num{64.1} respectively. 
  
We next present the results of dynamic measurements on micro-magnets. The AC susceptibility is measured by 
applying an excitation $\Delta H$ at frequency $f_{ac}$ and measuring  the change in the dipole moment 
$\Delta m$ of the material under study. The AC susceptibility $\chi_{ac}$ is defined as
\begin{equation}
    \chi_{ac} = \frac{1}{V} \frac{\Delta m}{\Delta H},
    \label{eq:magnetization}
\end{equation}
where $V$ is the volume of the micro-magnet.
The dynamic measurements are done by applying a single microwave frequency, at a frequency of maximum slope away from 
the resonance to the microwave antenna and measuring PL response from the diamond. The slope of a 
Lorentzian at this frequency point is equal to $A/0.77\Gamma$\,\cite{PhysRevB.84.195204}, where $A$ is the 
amplitude of the Lorentzian curve and $\Gamma$ is the 
full-width at half-maximum of the Lorentzian curve. Small shifts in resonance from pixel to pixel due to 
small changes in stray magnetic fields $\Delta 
B_{NV}$ manifest as changes in PL intensity $\Delta \mathrm{PL}$ as 
\begin{equation}
   \Delta \mathrm{PL} =  \frac{A}{0.77\Gamma} \times  \gamma_e \Delta B_{NV}.
   \label{eq:single_point}
\end{equation}
This 
makes it possible to reconstruct the magnetic field by recording the PL at a single microwave 
frequency. However, this method poses a challenge while measuring the dynamic response of magnetic 
materials to changes in applied magnetic fields. Due to significant shifts in resonance frequency owing to large changes in the applied magnetic field, the PL response is no longer linear to changes in the magnetic field. Hence, Eq.\,\eqref{eq:single_point} is no longer valid. To overcome this 
problem, we attach a current sensor of \SI{1.2}{\kilo \hertz} bandwidth to the current carrying coil 
(See Supplemental Information Section 1).    
The magnetic field from the current carrying coil is always proportional to the current, and the resonance
frequency of the \ce{NV-} center is a linear function of the magnetic field. Hence, the resonance frequency of 
the \ce{NV-} center is a a linear  function of the current in the coil. The output voltage of the current sensor 
is connected to the signal generator (SRS - SG380), which  functions in  a voltage-controlled 
oscillator mode. The output frequency of the function generator is given by
\begin{equation}
    f_o = f_c + v\times f_{dev},
\end{equation}
\begin{figure}[t]
  \centering
  \includegraphics[width=\linewidth]{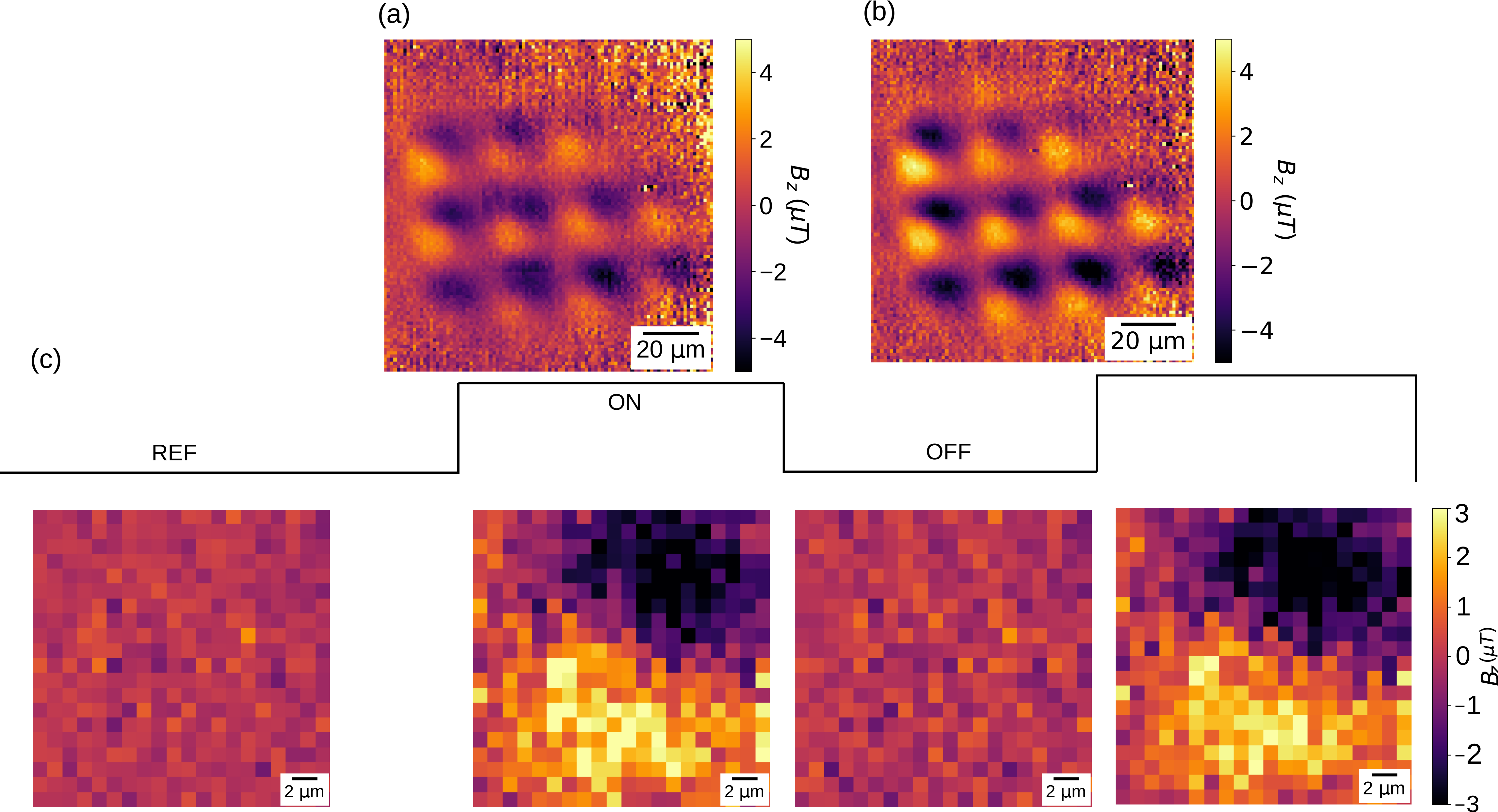}
  \caption{(a). Stray DC magnetic field maps obtained by fitting the full ODMR spectrum at applied fields of
  \SI{0.75}{\milli \tesla}, and \SI{1.85}{\milli \tesla}. (b). Change in stray field from micro-magnet \emph{A} when the magnetic field is varied between \SI{0.75}{\milli \tesla}, and \SI{1.85}{\milli \tesla}. Frame REF is  obtained by a \SI{0.75}{\milli \tesla} for elongated time and acts as reference.
  Subsequently, frames acquired when the excitation field is high are collectively labeled ON frames and 
  those with low excitation field are collectively named OFF frames. }
  \label{fig:time}
\end{figure} 
where $f_o$ is the output frequency, $f_c$ is the center frequency, $f_{dev}$ is frequency deviation and
$v$ is the applied voltage. The center frequency $f_c$, and the frequency deviation $f_{dev}$ can be 
set through software. Hence, by carefully choosing the values of $f_c$, and $f_{dev}$,  one can always ensure that the applied micro-wave frequency is  set to a point
where the slope and hence the sensitivity is maximum. This kind of a 
closed loop control ensures that the
the change in PL, $\Delta \mathrm{PL}$ given in Eq.\,\eqref{eq:single_point} always arises from the 
response of magnetic material under investigation and not the excitation field itself. To perform AC 
susceptibility measurements we chose two currents in the current carrying coil -  
\SI{1.45}{\ampere}, and \SI{3.6}{\ampere}, which correspond to an in-plane magnetic fields of 
\SI{0.75}{\milli \tesla} and \SI{1.85}{\milli \tesla} and a total magnetic field of \SI{0.97}{\milli 
\tesla} and \SI{2.40}{\milli \tesla}. The DC stray field images at both of these magnetic fields is
shown in Fig.\,\ref{fig:time}(a).  We then oscillate the magnetic field at a rate of \SI{0.5}{\hertz} between these two values of
magnetic fields and measure the PL at a rate of 1 fps. Each frame is obtained by averaging the PL for 
\SI{1}{\second}. First few frames are obtained at the lower magnetic field \SI{0.85}{\milli \tesla} for reference. The acquisition of camera frames is synchronized with the applied magnetic field. Therefore, 
alternate frames contain PL maps obtained for lower value of magnetic field and higher value of magnetic 
field respectively. Alternate frames are then subtracted from each other and scaled according to 
Eq.\,\eqref{eq:single_point}. Note that the amplitude $A$ and line-width $\Gamma$  given in 
Eq.\,\eqref{eq:single_point} is not 
uniform throughout the imaging window, and hence the scaling has to be done pixel by pixel. In Fig.\,\ref{fig:time}(a), magnetic field images are shown frame by frame after 
subtracting the reference frame REF while the magnetic field is low. The subtracted frames are indicative of the
change in stray magnetic field produced by the micro-magnets due to the change in dipole moment. 
The change in stray magnetic field is directly proportional to the change in dipole moment $\Delta m$ of 
Eq.\,\eqref{eq:magnetization}. According to the cylindrical model, the change in stray field    
corresponds to a   
change of \SI[parse-numbers = false]{39.6}{\femto \joule \per \tesla} in  dipole moment.  The maximum change in  stray field
produced by the micro-magnets is \SI[parse-numbers = false]{2.9\pm 0.2}{\micro \tesla}, while in other pixels it is less than that.
Note that the sensitivity is for magnetic fields along the direction of \ce{NV-} axis, while it will be less
for fields that are not along the \ce{NV-} axis. 
Given
our mean per pixel sensitivity of \SI{0.9}{\micro \tesla \per \sqrt{\hertz}}\,\cite{Parashar2022}, we still have to average more cycles to get decent SNR. The volume normalized susceptibility is found to be approximately same as the DC susceptibility $\chi_{ac,0.5} = 73.1, 61, 66.5$ 
for micromagnets \emph{A}, \emph{B}, \emph{C}. 
In Fig.\,\ref{fig:ac}, we show the results for higher frequencies.  
At higher frequencies, it also important to measure the imaginary part of AC susceptibility $\chi''_{ac}$. 
To facilitate the measurement of $\chi''_{ac}$, we acquire several frames at the lower magnetic field before 
applying the AC excitation as shown in Fig.\,\ref{fig:time}(b) (labeled as REF). After application of AC 
excitation, the imaginary part of AC susceptibility is proportional to REF-OFF, $\chi'' \propto $ REF-OFF, 
and the real part is proportional to ON-OFF, $\chi' \propto $ ON-OFF. 
In Fig.\,\ref{fig:ac}(a), the difference in stray magnetic field ON-OFF and in Fig.\,\ref{fig:ac}(b) OFF-REF is shown at a rate 
\SI{18}{\hertz} for the entire array. At this frequency, we observe no out of phase component 
in dipole moment as is expected from \ce{Py}\,\cite{BOOTH2021167631}. 
Finally in Fig.\,\ref{fig:ac}(c), the
AC susceptibility at various frequencies for micro-magnets $A,B,C$ are shown. The AC susceptibility for all the
frequencies measured are nearly equal to  DC susceptibility - $\chi_A=$\num[parse-numbers = false]{73 \pm 5}, $\chi_B=$\num[parse-numbers = false]{62 \pm 4}, $\chi_B=$\num[parse-numbers = false]{67 \pm 4}. The maximum frequency  of \SI{18}{\hertz} 
that we applied in our case was limited by software synchronization  pulse that we applied, to synchronize
the AC magnetic field and camera acquisition, and the frequency sweep limit of our signal generator, which 
is \SI{120}{\hertz}.  However, without these limitations in place, the higher frequency limit is 
dictated 
by the \ce{NV-} response time. The fastest response time reported has been around \SI{50}{\kilo 
\hertz}\,\cite{PhysRevLett.106.030802}. Assuming that at-least \num{4} demodulation cycles are required for
decent SNR, the maximum frequency at which one can operate will be around \SI{10}{\kilo \hertz} to \SI{15}{\kilo \hertz}. 

\section{Conclusion}
In conclusion, we propose and demonstrate a method to perform microscopic AC susceptibility measurements
using a quantum diamond microscope. We utilized a fast lock-in camera and a secondary sensor to 
dynamically vary the microwave frequency to measure AC susceptibilities of soft \ce{Py} micro-magnets.
Our work offers quantitative AC characterization over a wide-field of view in contrast to existing state of
art spatially resolved AC susceptibility measurement techniques\cite{doi:10.1063/1.3360205}. In particular our work can extend the application of quantum diamond microscope to simultaneously provide quantitative
information of magnetic properties as well as relaxation times of biologically relevant individual and agglomerates of magnetic nano-particles\,\cite{Zhang2021,https://doi.org/10.1002/pssa.201800254,Sandler2019}. Overall, our study is significant to physical, biological, and materials sciences and can be used in quantitative dynamic microscopic analyses of magnetic materials.

\begin{figure}[t]
  \centering
  \includegraphics[width=1\linewidth]{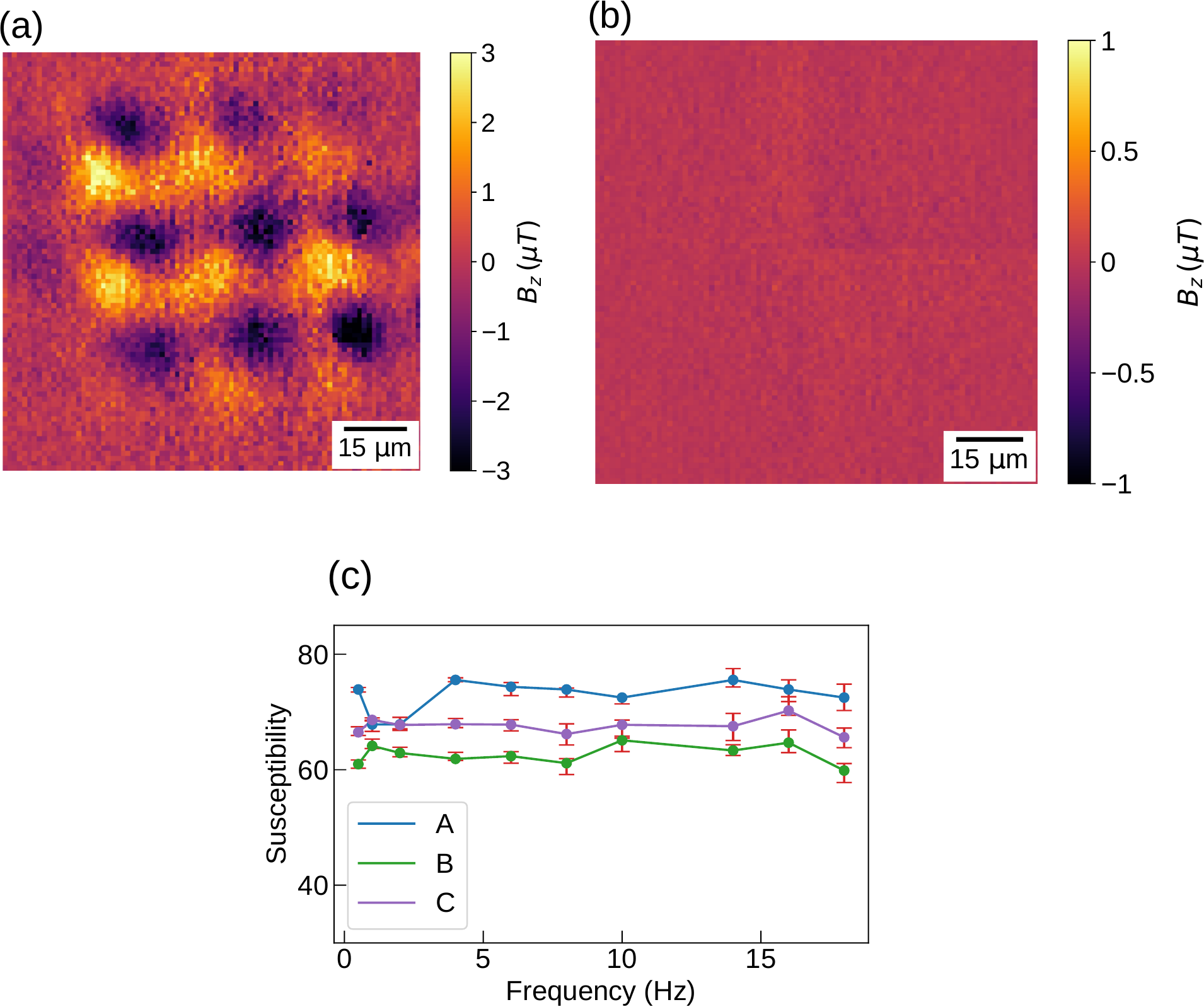}
  \caption{(a). In-phase and out-of-phase stray magnetic field from the micro-magnet array at a frequency of \SI{18}{\hertz} obtained by subtracting frames ON and OFF and OFF and REF respectively as shown in Fig.\,\ref{fig:time}.(b). Real part of  susceptibility  ($\chi'$) of micro-magnet \emph{A}, \emph{B}, and \emph{C} plotted against frequency applied excitation frequency.}
  \label{fig:ac}
\end{figure} 

\section{Author Contribution}
D.S., M.P. and K.S. conceived the idea. DS fabricated and characterized the devices, performed all experiments and analysed the data. MP and KS built the 
initial experimental setup. DS wrote the manuscript in discussion with all authors. KS supervised all 
aspects of the work.

\begin{acknowledgement}
K.S. acknowledges financial support from DST Inspire Faculty Fellowship \\
DST/INSPIRE/04/2016/002284, SERB EMR 
grant Number EMR/2016/007420, Asian Office of Aerospace Research and Development (AOARD) R\&D Grant no. 
FA2386-19-1-4042 and DST Quest Grant DST/ICPS/QuST/Theme-2/2019/Q-58. K.S. acknowledges the support and 
usage of fabrication facilities in the IIT Bombay Nano-fabrication facility via the NNetra project sponsored 
by Department of Science and Technology (DST) and Ministry of Electronics and Information Technology (MEITY),
India. The authors acknowledge insightful discussions with Prof. Anjan Barman (S. N. Bose National Center for Basic Sciences). K.S. also acknowledges useful discussions with Dr. Shamashis Sengupta (CNRS, Université Paris-Saclay) and Prof. Sumiran Pujari (IIT Bombay).
\end{acknowledgement}
\providecommand{\latin}[1]{#1}
\makeatletter
\providecommand{\doi}
  {\begingroup\let\do\@makeother\dospecials
  \catcode`\{=1 \catcode`\}=2 \doi@aux}
\providecommand{\doi@aux}[1]{\endgroup\texttt{#1}}
\makeatother
\providecommand*\mcitethebibliography{\thebibliography}
\csname @ifundefined\endcsname{endmcitethebibliography}
  {\let\endmcitethebibliography\endthebibliography}{}

\end{document}